\documentclass[12pt]{article}
\usepackage{fullpage}

\def\nofigures{0} 

\def\dccsub{0}

\usepackage{graphicx,amsfonts,amsmath,amssymb,epsfig,hyperref,times,color}
\usepackage[ruled, noline]{algorithm2e}
\usepackage{fancybox}
\usepackage{hyperref}
\usepackage{setspace}
\usepackage{pgfplots}
\pgfplotsset{compat=newest}
\ifnum\nofigures=0
\usepackage{tikz}
\fi

\newcommand{\mnote}[1]{ \marginpar{\tiny\bf
            \begin{minipage}[t]{0.5in}
              \raggedright #1
           \end{minipage}}}

\renewcommand{\mnote}[1]{}

\newcommand{\fmnote}[1]{}

\newtheorem{thm}{Theorem}
\newcommand{\BT}{\begin{thm}} \newcommand{\ET}{\end{thm}}
\def\FullBox{\hbox{\vrule width 8pt height 8pt depth 0pt}}
\newcommand{\qed}{\;\;\;\FullBox}
\newcommand{\bsize}{\lfloor \log n \rfloor}
\newenvironment{proof}{\noindent{\bf Proof:~~}}{\(\qed\)}
\newcommand{\BPF}{\begin{proof}} \newcommand {\EPF}{\end{proof}}
\newenvironment{proofof}[1]{\noindent{\bf Proof of {#1}:~~}}{\(\qed\)}
\newcommand{\BPFOF}{\begin{proofof}} \newcommand {\EPFOF}{\end{proofof}}
\newcommand{\qedsketch}{\;\;\;\Box}

\newenvironment{smallproof}{\noindent{\bf Proof:~~}}{\(\qedsketch\)}
\newcommand{\bpf}{\begin{smallproof}} \newcommand{\epf}{\end{smallproof}}
\newcommand{\eqdef}{\stackrel{\rm def}{=}}
\newcommand{\BE}{\begin{enumerate}} \newcommand{\EE}{\end{enumerate}}
\newtheorem{lem}{Lemma}      
\newcommand{\BL}{\begin{lem}} \newcommand{\EL}{\end{lem}}
\newtheorem{clm}[lem]{Claim}
\newcommand{\BCM}{\begin{clm}} \newcommand{\ECM}{\end{clm}}
\newcommand{\BI}{\begin{itemize}} \newcommand{\EI}{\end{itemize}}
\newtheorem{defn}{Definition}         
\newcommand{\BD}{\begin{defn}} \newcommand{\ED}{\end{defn}}
\newtheorem{cor}[thm]{Corollary}      
\newcommand{\BC}{\begin{cor}} \newcommand{\EC}{\end{cor}}
\newcommand{\eps}{\epsilon}
\newcommand{\poly}{{\rm poly}}
\newcommand{\bitset}{{\{0,1\}}}

\begin{document}

\date{}
\title{A simple online competitive adaptation of Lempel-Ziv compression
 with efficient random access support }

\author{Akashnil Dutta\thanks{CSAIL, MIT, Cambridge MA 02139.
E-mail: {\tt akashnil@mit.edu}.
}
\and
Reut Levi\thanks{School of Computer Science, Tel Aviv University.
E-mail: {\tt reuti.levi@gmail.com}.
} \and
Dana Ron \thanks {School of Electrical Engineering, Tel Aviv
University.
E-mail: {\tt danar@eng.tau.ac.il}.
}
\and
Ronitt Rubinfeld\thanks{CSAIL, MIT, Cambridge MA 02139 and
the Blavatnik School of Computer Science, Tel Aviv University.
 E-mail: {\tt  ronitt@csail.mit.edu}.
}
}
\maketitle

\begin{abstract}
We present a simple adaptation of the Lempel Ziv 78' (LZ78)
compression scheme 
\ifnum\dccsub=0
({\em IEEE Transactions on Information Theory, 1978\/})
\fi
that supports efficient random access to the input string.
\ifnum\dccsub=0
Namely, given query access to the compressed string,
it is possible to efficiently recover any symbol
of the input string. 
\fi 
The compression algorithm is given as input a
parameter $\eps >0$, and with very high probability increases the
length of the compressed string by at most a factor of $(1+\eps)$.
The access time is $O(\log n + 1/\eps^2)$
in expectation, and $O(\log n/\eps^2)$
with high probability.
The scheme relies on sparse transitive-closure spanners.
Any (consecutive) substring
of the input string can be retrieved at an additional additive cost
in the running time of the length of the substring.
\ifnum\dccsub=0
We also formally establish the necessity of modifying LZ78 so as
to allow efficient random access. Specifically,
we construct a family of strings for which $\Omega(n/\log n)$
queries  to the LZ78-compressed string are required in order to
recover
a single symbol in the input string.
\fi
The main benefit of the proposed scheme is that it preserves
the online nature and simplicity of LZ78, and that for
{\em every\/} input string, the length of the compressed string is
only a small factor larger than that obtained by running LZ78.
\end{abstract}

\section{Introduction}
As the sizes of our data
sets are skyrocketing it is become important to allow a user to  
access any desired {\em portion} of
the original data without
decompressing the entire dataset. This problem has been
receiving quite a bit of recent attention
(see e.g.~\cite{SG,BLRSSW,FV07,KN,DPT10,FN09,CSW09}).
Compression and decompression schemes that
allow fast random-access decompression  support have
been proposed with the aim
of achieving similar compression rates to the known and widely used
compression schemes, such as arithmetic coding~\cite{WNC87},
LZ78~\cite{LZ78}, LZ77~\cite{LZ77} and Huffman coding~\cite{Huffman}.

In this work, we focus on adapting the widely used
LZ78 compression scheme
so as to allow fast random access support.
Namely, given access to the compressed string and
a location $\ell$ in the original uncompressed string, we would like
 to be able to efficiently recover the $\ell$-th symbol in the uncompressed
string. More generally, the goal is to efficiently
recover a substring starting at location $\ell_1$ and
ending at location $\ell_2$ in the uncompressed string.
Previously, Lempel Ziv-based schemes were designed to
support fast random access, in particular,
based on LZ78~\cite{SG}, LZ77~\cite{KN} and
as a special case of grammar-based compression~\cite{BLRSSW}.

The first basic question that one may ask is whether there is
any need at all to modify the LZ78 scheme in order to support
fast random access. 
\ifnum\dccsub=0
We formalize the intuition that this is
indeed necessary and show that without any modifications
every (possibly randomized) algorithm will need time linear in
the length of the LZ78-compressed string to recover a single
symbol of the uncompressed string.
\else
In our extended version~\cite{DLRR-long} we show that
any algorithm (even randomized) needs at least linear time in the length
of the LZ78 compressed string to recover a single symbol.
\fi

Having established that some modification is necessary,
the next question is how do we evaluate the
compression performance of a compression
scheme that is a modification of LZ78
and supports efficient random access.
As different strings have very different compressibility
properties according to LZ78,
in order to compare the quality of a new scheme to LZ78, we
consider
a competitive analysis framework.
In this framework,
we require that for every input string, the
length of the compressed string is a most
multiplicative factor of $\alpha$ larger than the
length of the LZ78-compressed string,
 where $\alpha > 1$ is a small constant.
For a randomized compression algorithm this should hold
with high probability (that is, probability $1-1/\poly(n)$ where
$n$ is the length of the input string).
If this bound holds (for all strings) then
we say that the scheme is $\alpha$-{\em competitive} with LZ78.

One
additional feature of interest is whether the modified
compression algorithm preserves the online nature of LZ78.
The LZ78 compression algorithm works by outputting a sequence
of {\em codewords\/}, where each codeword encodes a
(consecutive)
substring of the input string, referred to as a {\em phrase\/}.
LZ78 is online in the sense that if the compression
algorithm is stopped at any point,
then we
can
recover all phrases encoded by
the codewords output until that point.
Our scheme preserves this property of LZ78 and furthermore,
supports online random access. Namely, at each point
in the execution of the
compression algorithm we can efficiently recover
any symbol (substring) of the input string
 that has already been encoded.
A motivating example to keep in mind is of a powerful server that receives a stream of data over a long period of time. All through this period of time the server sends the compressed data to clients which can, in the meantime, retrieve portions of the data efficiently. This scenario fits cases where the data is growing incrementally, as in log files or user-generated content.

\ifnum\dccsub=0
\vspace{-1.5ex}
\fi
\subsection{Our Results}\label{subsec:our-res}
\ifnum\dccsub=0
We first provide a deterministic compression algorithm
which is $3$-competitive with LZ78 (as defined above),
and a matching random access algorithm which
runs in time $O(\log n)$, where $n$ is the length of
the input string. This algorithm retrieves any requested
single symbol of the uncompressed string.
By slightly adapting this
algorithm it is possible to
retrieve a substring of length $s$ in time
$O(\log n) + s$.

Thereafter, we provide a randomized compression algorithm
which for any chosen epsilon is $(1+\eps)$-competitive with LZ78.
\else
We provide a randomized compression algorithm
which for any chosen epsilon is $(1+\eps)$-competitive with LZ78.
\fi
The expected running time of the matching random access
algorithm is $O(\log n + 1/\epsilon^{2})$,
and with high probability
is bounded by
\ifnum\dccsub=0
\footnote{This bound can be improved to
$O((\log n/\eps + 1/\eps^2)\log(\log n/\eps))$, but this
improvement comes at a cost of making the algorithm somewhat more
complicated, and hence we have chosen only to sketch this improvement (see Subsection~\ref{sec:imp-run}).}
\else
\footnote{This bound can be improved to
$O((\log n/\eps + 1/\eps^2)\log(\log n/\eps))$ with a somewhat more
complicated algorithm (see our extended version~\cite{DLRR-long}).}
\fi
$O(\log n/\eps^2)$. The probability
is taken over the random coins of the randomized compression algorithm.
\ifnum\dccsub=0
As before, a substring can be recovered in time
that is the sum of the (single symbol)
random access time and the length of the string.
\else
A substring can be recovered in time
that is the sum of the (single symbol)
random access time and the length of the string.
\fi
Similarly to LZ78, the scheme works in an online manner in the sense
described above. 
The scheme is fairly simple and does not require any sophisticated data structures.
\ifnum\dccsub=0
For the sake of simplicity we describe them for the case in which
the alphabet of the input string is $\bitset$, but they can
easily be extended to work for any alphabet $\Sigma$.

As noted previously, we also give a lower bound that is linear in the
length of the compressed string for any random access algorithm
that works with (unmodified) LZ78 compressed strings.
\fi
\ifnum\dccsub=1

{\bf Experimental Results.}
\else
\paragraph{Experimental Results.}
\fi
We provide experimental results which demonstrate that our scheme is competitive and that random access is extremely efficient in practice. An implementation of
our randomized scheme is available online~\cite{DLRR}.
\ifnum\dccsub=1
\vspace{-2ex}
\fi
\subsection{Techniques}\label{subsec:techniques}
The LZ78 compression algorithm outputs
a sequence of codewords, each encoding a phrase (substring)
of the input string. Each phrase is the concatenation of
a previous phrase and one new symbol.
The codewords are constructed
sequentially, where each codeword consists
of an index $i$ of a previously encoded phrase
(the longest phrase that matches a prefix of the yet
uncompressed part of the input string),
and one new symbol.
Thus the codewords (phrases they encode)
 can be seen as forming a directed tree,
which is a trie,
with an edge pointing from each child to its parent.
Hence, if a node $v$ corresponds to a phrase
$s_1,\dots,s_t$, then for each $1 \leq j \leq t$,
there is an ancestor node of $v$ that corresponds
to the prefix $s_1,\dots,s_j$, and is encoded by the codeword $(i,s_j)$
(for some $i$), so that $s_j$ can be ``revealed''
by obtaining this codeword.

In order to support random access, we want to be able to
perform two tasks. 
The first task is to
identify, for any given index $\ell$, what is the codeword
that encodes the phrase to which the $\ell$-th symbol of
the input string belongs.
We refer to this codeword
as the ``target codeword''.
Let $p$ denote starting position of the corresponding phrase
(in the input string),
then the second task is to navigate
(quickly) up the tree (from the node corresponding
to the target codeword) and reach the ancestor node/codeword
at depth $\ell-p+1$ in the tree. This codeword reveals
the symbol we are looking for.
In order to be able to perform these two tasks efficiently,
we modify the LZ78 codewords.
To support the first task we add information concerning the
position of phrases in the input (uncompressed) string.
To support the second task we add additional pointers to
ancestor nodes in the tree, that is, indices of
encoded phrases that correspond to such nodes.
Thus we (virtually) construct a (very sparse)
Transitive Closure (TC) spanner~\cite{BGJRW} on the tree.
The spanner allow to navigate quickly between pairs of
codes.

When preprocessing is allowed, both tasks can be achieved more efficiently using auxiliary data structures. Specifically, the first task can be achieved using rank and select queries in time complexity $O(1)$  (see e.g.~\cite{GHSV07}) and the second task can be achieved in time complexity $O(\log \log n)$ via level-ancestor queries on the trie  (see e.g.~\cite{FM08}). However, these solutions are not adaptable, at least not in a straightforward way, to the online setting and furthermore the resulting scheme is not $(1+\eps)$-competitive with LZ78 for every $\eps$.  

\ifnum\dccsub=0
In the deterministic scheme, which is $3$-competitive with
LZ78, we include the additional information (of the position and
one additional pointer) in every codeword, thus making
it relatively easy to perform both tasks in time $O(\log n)$.
\fi
In order to obtain
the scheme that is $(1+\eps)$-competitive with LZ78 we include
the additional information only in an $O(\eps)$-fraction of
the codewords, and the performance of the tasks becomes
more challenging. Nonetheless, the dependence of the
running time on $n$ remains logarithmic (and the dependence on
 $1/\eps$ is polynomial).
\ifnum\dccsub=0

The codewords which include additional information are chosen randomly in order to spread them out evenly in the trie.
It is fairly easy to obtain similar results if the structure of the trie is known in advance, 
however, in an online setting, the straightforward deterministic approach can blow up the size of the output by a large factor.  
\fi
\ifnum\dccsub=1
\vspace{-1.5ex}
\fi
\subsection{Related Work}
\label{subsec:rel-work}
Sadakane and Grossi~\cite{SG} give a compression scheme
that supports the retrieval
of any $s$-long consecutive substring of
an input string $S$ of length $n$ over alphabet
$\Sigma$ in
$O(1+s/(\log_{|\Sigma|} n))$ time.
In particular, for a single symbol in the input string
the running time is $O(1)$. The number of bits in
the compressed string is upper bounded by
$nH_k(S) +
O\left(\frac{n}{\log_{|\Sigma|}n}\left(k\log |\Sigma|+\log\log n\right)\right)$,
where $H_k(S)$ is the $k$-th order empirical entropy of $S$.
Since their compression algorithm builds on LZ78,
the bound on the length of the compressed string
for any given input string can actually
be expressed as the sum of the length of the LZ78
compressed string plus $\Theta(n\log \log n/\log n)$ bits for supporting rank and select operations in constant time~\footnote{The $\Theta(n\log \log n/\log n)$ space requirement can be decreased if one is willing to spend more than constant time.}.
They build on the LZ78 scheme in the sense that they store suits
of data structures that encode the structure of the LZ78 trie
and support fast random access.
Hence, for input strings that are compressed by LZ78
to a number of bits that is at least on the order of
 $n\log\log n/\log n$, their result
is essentially the best possible as compared to LZ78.
However, their scheme is not in general competitive
(as defined above)
with LZ78 because of its performance on
highly compressible strings.
We also note that their compression algorithm
does not work in an online fashion, but rather constructs
all the supporting data structures given the complete
LZ78 trie.

Two alternative schemes which give the same space and time bounds
as in~\cite{SG} were provided by Gonz\'{a}lez and Navarro~\cite{GN}
and Ferragina and Venturini~\cite{FV07}, respectively.
They are simpler, where the first uses an arithmetic encoder and
the second does not use any compressor. (They also differ
in terms of whether $k$ has to be fixed in advance.)
By the above discussion the performance of these schemes is
not in general competitive with LZ78.

Kreft and Navarro~\cite{KN} provide a variant of LZ77 that
supports retrieval of any $s$-long consecutive substring of
$S$ in $O(s)$ time.
They show that in practice their scheme achieves close
results to LZ77 (in terms of the compression ratio).
However, the usage of a data structure that supports
the rank and select operations requires
 $\Omega(n\log \log n/\log n)$ bits.

The Lempel-Ziv compression family belongs to a wider family of schemes called
grammar-based compression schemes.
In these schemes the input string is represented by a context-free
grammar (CFG), which is unambiguous, namely, it generates a unique string.
Billie et al.~\cite{BLRSSW} show how to transform any grammar-based
compression scheme so as to support random access
in $O(\log n)$ time. The transformation increases the compressed
representation by a multiplicative factor (larger than $1$).

\section{Preliminaries}\label{sec:prel}
\ifnum\dccsub=1

{\bf The LZ78 compression scheme.}
\else
\paragraph{The LZ78 compression scheme.}
\fi
\ifnum\dccsub=0
Before we describe our adaptation of the LZ78 scheme~\cite{LZ78},
we describe the latter in detail.
The LZ78 compression algorithm receives an input string
$x\in\Sigma^n$ over alphabet $\Sigma$ and returns a list,
\fmnote{D: reconsider notation $C^x$}
$C^x = C^x_{\rm LZ}$,
of codewords of the form $(i, b)$, where $i\in \mathbb{N}$ and $b\in\Sigma$.
Henceforth, unless specified otherwise, $\Sigma = \{0,1\}$.
\fmnote{D: justify? or say that without loss of generality?}
Each codeword $(i, b)$ encodes a phrase, namely a substring of $x$,
which is the concatenation of the $i$-th phrase (encoded by
$C^x[i]$) and $b$, where we define the $0$-th phrase to be the empty string.
The first codeword is always of the form
$(0,x[1])$, indicating that the first phrase consists of a single
symbol $x[1]$. The compression algorithm continues scanning the input
string $x$ and partitioning it into phrases. When determining the
$j$-th phrase,
if the algorithm has already scanned $x[1,\dots,k]$, then the
algorithm finds the longest prefix
 $x[k+1,\dots,n-1]$ that is the same as a phrase with index  $i < j$.
 If this prefix is $x[k+1,\dots, t]$, then the algorithm outputs
 the codeword $(i,x_{t+1})$ (if the prefix is empty, then $i=0$).

An efficient (linear in $n$) LZ78 compression algorithm can be
implemented by maintaining an auxiliary trie (as illustrated in
Figure~\ref{fig:lz}, Section~\ref{fig:sec}).
The trie structure is implicit in the output of the LZ78 algorithm.
Namely, for an input string $x\in\{0,1\}^n$, the
trie $T^x=(V^x,E^x)$ is defined as follows.
For each codeword $C^x[i]$, $1 \leq i \leq m$ there is a node $v_i$ in $V^x$,
and there is also a node $v_0$ corresponding to the root of the tree.
If $C^x[j] = (i,b)$, then there is an edge between $v_j$ and $v_i$
(so that $v_i$ is the parent of $v_j$). Given the correspondence
between codewords and nodes in the trie, we shall sometimes
refer to them interchangeably.

\sloppy
In the course of the compression process, when constructing
the $j$-th codeword (after scanning $x[1,\dots,k]$) the compression
algorithm can find the longest prefix of $x[k+1,\dots,n-1]$
that matches an existing phrase $i$ simply by walking down the
trie. Once the longest match is found (the deepest node is reached),
a new node is added to the trie.
Thus the trie structure may be an actual data structure used
in the compression process, but it is also implicit in the
compressed string (where we think of a codeword $C^x[j] = (i,b)$ as
having a {\em pointer\/} to its {\em parent\/} $C^x[i]$).
Decompression can also be implemented in linear time by
iteratively recovering the phrases that correspond to the codewords
and essentially rebuilding the trie (either explicitly or implicitly).
In what follows, we refer to $i$ as the {\em index} of $C^x[i]$ and to $x[j]$ as the bit at {\em position} $j$.
\else
The LZ78 compression algorithm~\cite{LZ78} receives an input string
$x\in\Sigma^n$ over alphabet $\Sigma$ and returns a list,
\fmnote{D: reconsider notation $C^x$}
$C^x = C^x_{\rm LZ}$,
of codewords of the form $(i, b)$, where $i\in \mathbb{N}$ and $b\in\Sigma$.
Henceforth, unless specified otherwise, $\Sigma = \{0,1\}$.
\fmnote{D: justify? or say that without loss of generality?}
Each codeword $(i, b)$ encodes a phrase, namely a substring of $x$,
which is the concatenation of the $i$-th phrase (encoded by
$C^x[i]$) and $b$, where we define the $0$-th phrase to be the empty string.
An efficient (linear in $n$) LZ78 compression algorithm can be
implemented by maintaining an auxiliary trie.
The trie structure is implicit in the output of the LZ78 algorithm.
Namely, for an input string $x\in\{0,1\}^n$, the
trie $T^x=(V^x,E^x)$ is defined as follows.
For each codeword $C^x[i]$, $1 \leq i \leq m$ there is a node $v_i$ in $V^x$,
and there is also a node $v_0$ corresponding to the root of the tree.
If $C^x[j] = (i,b)$, then there is an edge between $v_j$ and $v_i$
(so that $v_i$ is the parent of $v_j$). Given the correspondence
between codewords and nodes in the trie, we shall sometimes
refer to them interchangeably.
In what follows, we refer to $i$ as the {\em index} of $C^x[i]$ and to $x[j]$ as the bit at {\em position} $j$.
\fi
\ifnum\dccsub=1

{\bf Competitive schemes with random access support.}
\else
\paragraph{Competitive schemes with random access support.}
\fi
We aim to provide a scheme, $\mathcal{A}$, which compresses every input string
almost as well as LZ78 and supports efficient local decompression.
Namely, given access to a string that is the output of $\mathcal{A}$
on input $x$
 and $1\leq \ell_1 \leq \ell_2 \leq n$, the
local decompression algorithm outputs $x[\ell_1, \ldots, \ell_2]$
efficiently. In particular, it does so without decompressing the entire string.
We first describe our scheme for the case where $\ell_1 = \ell_2$,
which we refer to as {\em random access\/}, and later explain how to extend
the scheme for $\ell_1 < \ell_2$.
The quality of the compression is measured with respect to LZ78,
formally, we require the scheme to be {\em competitive} with LZ78
as defined next. We note that here and in all that follows, when
we say ``with high probability'' we mean with probability at least
$1-1/\poly(n)$.
\ifnum\dccsub=1
\vspace{-1.5ex}
\fi
\BD[Competitive schemes]\label{def:competitive}
Given a pair of deterministic compression algorithms
$\mathcal{A}:\{0,1\}^*\rightarrow\{0,1\}^*$ and
$\mathcal{B}:\{0,1\}^*\rightarrow\{0,1\}^*$,
we say that algorithm $\mathcal{B}$ is $\alpha$-competitive with $\mathcal{A}$
if for every input string $x\in\{0,1\}^*$,
we have $|C_\mathcal{B}^x| \leq \alpha|C_\mathcal{A}^x|$,
where $C_\mathcal{B}^x$ and $C_\mathcal{A}^x$ are the compressed
strings output by $\mathcal{A}$ and $\mathcal{B}$, respectively,
on input $x$.
For a randomized algorithm $\mathcal{B}$, the requirement is that
$|C_\mathcal{B}^x| \leq \alpha|C_\mathcal{A}^x|$ with high probability
over the random coins of $\mathcal{B}$.
\ED

\ifnum\dccsub=1
\vspace{-1.5ex}

{\bf Word RAM model.}
\else
\paragraph{Word RAM model.}
\fi
We consider the RAM model with word size $\log n +1$, where $n$ is the
length of the input string.\ifnum\dccsub=1
\footnote{If instead of $\log n +1$ we take the word size to be $\log m + 1$, where $m$ is the number of phrases in the compressed string, then our results remain effectively the same.}
\fi 
We note that it suffices to have an upper bound on this value
in order to have a bound on the number of bits for representing any index of a
phrase.
A codeword of LZ78 is one word, i.e., $i$ and $b$ appear
consecutively where $i$ is represented by $\log n$ bits.
\ifnum\dccsub=0
For the sake of clarity of the presentation, we write it as $(i,b)$.
Our algorithms \mnote{D: can reconsider last par.}
(which supports random access) use words of
size $\log n +1$ as well. If one wants to consider variants of LZ78
that apply
bit optimization and/or work when an upper bound on the length of the
input string is not known in advance, then our algorithms need to
be modified accordingly so as to remain competitive (with the
same competitive ratio).
\fi

\ifnum\dccsub=0
We wish to point out that if we take the word size to be $\log m + 1$ (instead of $\log n + 1$), where $m$ is the number of phrases in the compressed string, then our results remain effectively the same. Specifically, in the worst case the blow up in the deterministic scheme is of factor of $4$ (instead of $3$) and in the randomized scheme is of factor $(1+2\eps)$ (instead of ($1+\eps$)).
\fi
\ifnum\dccsub=1
\vspace{-1.5ex}
\fi
\ifnum\dccsub=0
\section{A Deterministic Scheme}\label{sec:det}
In this section we describe a simple
deterministic compression scheme
which is based on the LZ78 scheme.
\else
\section{A Randomized Scheme}
In this section we first describe a simple
deterministic compression scheme
which is based on the LZ78 scheme.
Thereafter, we present our randomized compression scheme which
builds on this deterministic scheme.
\fi

In the deterministic compression scheme, to each codeword we add a pair of additional entries.
The first additional entry is the starting position of the
encoded phrase in the uncompressed string.
On an input $x\in\{0,1\}^n$ and $1\leq \ell \leq n$,
this allows the algorithm to efficiently find the codeword encoding
the phrase that contains the $\ell$-th bit by performing
a binary search on the position entries.
The second entry we add is an extra pointer (we shall use the
terms ``pointer'' and ``index'' interchangeably).
Namely, while in LZ78 each codeword indicates the index of the
former codeword, i.e., the direct parent in the trie,
(see Section~\ref{sec:prel}), we add another index, to an
ancestor node/codeword (which is not the direct parent).
In order to allow efficient random access, our goal is to
guarantee that for every pair of connected nodes,
$u, v$ there is a short path connecting $u$ and $v$.
Namely, if we let $d_G(u, v)$ denote the length of the shortest
path from $u$ to $v$ in a directed graph $G$, then the requirement
is that for $u,v$ such that $d_G(u, v) < \infty$ it holds that
$d_G(u, v)$ is small.
Before we describe how to achieve this property on
(a super-graph of) the constructed trie
we describe how to guarantee the property on a simple directed path.
Formally we are interested in constructing a Transitive-Closure (TC)
spanner, defined as follows:
\ifnum\dccsub=1
\vspace{-1.5ex}
\fi
\BD[TC-spanner~\cite{BGJRW}]\label{def:span}
Given a directed graph $G = (V,E)$ and an integer $k\geq 1$,
a $k$-transitive-closure-spanner ($k$-TC-spanner) of $G$
is a directed graph $H = (V, E_H)$ with the following properties:
\ifnum\dccsub=0
\BE
\item $E_H$ is a subset of the edges in the
transitive closure\footnote{The transitive closure of a graph $G=(V,E)$ is the
graph $H=(V',E')$ where $V' = V$ and $E' = \{(u,v): d_G(u,v) < \infty\}$.}
of $G$.
\item For all vertices $u,v \in V$, if $d_G(u,v) < \infty$,
then $d_H(u,v) \leq k$.
\EE
\else
(1) $E_H$ is a subset of the edges in the
transitive closure\footnote{The transitive closure of a graph $G=(V,E)$ is the
graph $H=(V',E')$ where $V' = V$ and $E' = \{(u,v): d_G(u,v) < \infty\}$.}
of $G$.
(2) For all vertices $u,v \in V$, if $d_G(u,v) < \infty$,
then $d_H(u,v) \leq k$.
\fi
\ED

\ifnum\dccsub=0
\subsection{TC Spanners for Paths and Trees}
\else
\subsection{A Randomized Algorithm via TC-Spanners}
\fi
Let $\mathcal{L}_n = (V,E)$ denote the directed line (path)
over $n$ nodes (where edges are directed ``backward'').
Namely, $V=\{0,\ldots,n-1\}$ and
$E = \{(i,i-1) : 1 \leq i \leq n-1\}$.
Let $f_n(i) \; {\buildrel\rm def\over=} \; i\mod \bsize$ and
let $E' = \{(i,\max\{i-2^{f_n(i)}\cdot \bsize, 0\}) : 1\leq i\leq n-1 \}$.
Observe that each node $1 \leq i \leq n-1$ has exactly
one outgoing edge in $E'$ (in addition to the single outgoing edge in $E$).
Define $\mathcal{H}_n = (V,E\cup E')$.
\fmnote{D: maybe can save a $\log n$ terms by $-1$}
\BCM\label{clm:span}
$\mathcal{H}_n$ is  a
$(4\log n)$-TC-spanner
of $\mathcal{L}_n$.
\ECM
\ifnum\dccsub=1
\vspace{-.5ex}
\BPF
For every $0 \leq r < t \leq n-1$, consider the following algorithm to
get from $t$ to $r$ (at each step of the algorithm stop if $r$ is reached):
(1)\label{st:span1} Starting from $t$ and using the edges of
$E$, go to the first node $u$ such that $f_n(u) = \bsize -1$.
(2)\label{st:span2} From $u$ iteratively proceed by taking the
outgoing edge in $E'$ if it
does not go beyond $r$
(i.e., if the node reached after taking the edge is not smaller than $r$),
and taking the outgoing edge in $E$ otherwise. See~\cite{DLRR-long} for proof of correctness.
\EPF
\else
\BPF
For every $0 \leq r < t \leq n-1$, consider the following algorithm to
get from $t$ to $r$ (at each step of the algorithm stop if $r$ is reached):
\BE
\item\label{st:span1} Starting from $t$ and using the edges of
$E$, go to the first node $u$ such that $f_n(u) = \bsize -1$.
\item\label{st:span2} From $u$ iteratively proceed by taking the
outgoing edge in $E'$ if it
does not go beyond $r$
(i.e., if the node reached after taking the edge is not smaller than $r$),
and taking the outgoing edge in $E$ otherwise.
\EE
Clearly, when the algorithm terminates, $r$ is reached.
Therefore, it remains to show that the length of the path taken by the algorithm is bounded by $4\log n$.
Let $a(i)$ denote the node reached by the algorithm after taking $i$ edges in $E$ starting from $u$.
Therefore, $a(0) = u$ and $f_n(a(i)) =  \bsize -1 - i$ for every $0 \leq i < \bsize$ and $i \leq s$, where $s$ denotes the total number of edges taken in $E$ starting from $u$.
For every pair of nodes $w \geq q$ define $g(w,q) = \lfloor(w-q)/\bsize\rfloor$,
i.e., the number of complete blocks between $w$ and $q$.
Thus, $g(a(i), r)$ is monotonically decreasing in $i$, for $i \leq s$.
Consider the bit representation of $g(a(i), r)$.
If from node $a(i)$ the algorithm does not take the edge in $E'$ it is implied that the $j$-th bit in $g(a(i), r)$ is $0$ for every $j \geq f_n(a(i))$.
On the other hand, if from node $a(i)$ the algorithm takes the edge in $E'$ then after taking this edge the $f_n(a(i))$-th bit turns $0$.
Therefore by an inductive argument, when the algorithm reaches
$a(i)$, $g(a(i), r)$ is $0$ for every $j > f_n(a(i))$.
Thus, $g(a(\min\{\bsize -1, s\}), r) = 0$, implying that the total number of edges taken on $E'$ is at most $\log n$.
Combined with the fact that the total number of edges taken on $E$ in Step~\ref{st:span2} is bounded by $2\log n$ and the fact that the total number of edges taken on $E$ in Step~\ref{st:span1} is bounded by $\log n$, the claim follows.
\EPF
\fi

\medskip
\fmnote{D: ref to something similar?}
From Claim~\ref{clm:span} it follows that for every
$m < n$, $V=\{0,\ldots,m\}$, $E = \{(i,i-1) : 1 \leq i \leq m-1\}$
and $E' = \{(i,\max\{i-2^{f_n(i)}\cdot \bsize, 0\})$,
$(V,E\cup E')$ is a
$(4\log n)$-TC-spanner
of $\mathcal{L}_m$.
This implies a construction of a
$(4\log n)$-TC-spanner
for any tree on $n$ nodes.
Specifically, we consider trees where the direction of the
edges is from child to parent (as defined implicitly by
the codewords of LZ78)
and let $d(v)$ denoted the depth of a node $v$ in the tree
(where the depth of the root is $0$).
If in addition to the pointer to the parent, each node, $v$,
points to the ancestor at distance $2^{f_n(d(v))} \cdot \bsize$
(if such a node exists),
then for every pair of nodes $u, v$ on a path from a leaf
to the root, there is a path of length at most
$4\log n$
connecting $u$ and $v$.
\ifnum\dccsub=0

\smallskip
We note that using $k$-TC-spanners with $k = o(\log n)$ will not
improve the running time of our random access algorithms asymptotically
(since they perform an initial stage of a binary search).
\else
\footnote{We note that using $k$-TC-spanners with $k = o(\log n)$ will not
improve the running time of our random access algorithms asymptotically
(since they perform an initial stage of a binary search).}
\fi

\ifnum\dccsub=0
\subsection{Compression and Random Access Algorithms}
As stated at the start of this section,
in order to support efficient random access we modify the codewords of LZ78.
Recall that in LZ78 the codewords have the form $(i,b)$, where $i$ is the index
of the parent codeword (node in the trie) and $b$ is the additional bit.
In the modified scheme, codewords are of of the form
$W=(p,i,k,b)$,
where $i$ and $b$ remain the same, $p$ is the starting position
of the encoded phrase in the uncompressed string and $k$ is an index of
an ancestor codeword
(i.e., encoding a phrase that is a prefix of the phrase encoded by $W$).
As in  LZ78, our compression algorithm
(whose pseudo-code appears in
Algorithm~\ref{alg:com}, Subsection~\ref{detps.subsec})
maintains
a trie $\mathcal{T}$ as a data structure
where the nodes of the trie correspond to codewords
encoding phrases (see Section~\ref{sec:prel}).
Initially, $\mathcal{T}$ consists of a single root node.
Thereafter, the input string is scanned and a node is added to
the trie for each codeword that the algorithm outputs,
giving the ability
to efficiently construct the next codewords.
The data structure used is standard: for
each node the algorithm maintains the index of the phrase that corresponds
to it, its depth, and pointers to its children.

\fmnote{D: add notation for $C[t]$ and $C[r]$, e.g.,
$t= \tau(x,\ell)$ and $r= \rho(x,\ell)$, or don't use enough?}
Given access to a compressed string, which is a list of
codewords $C[1,\dots,m]$, and
an index $1 \leq \ell \leq n$,
the random access algorithm
(whose pseudo-code appears in
Algorithm~\ref{alg:local}, Subsection~\ref{detps.subsec})
first performs a binary search (using the position entries in
the codewords) in order to find the codeword, $C[t]$,
which encodes the phrase $x[\ell_1,\dots,\ell_2]$
containing the $\ell$-th bit of the input string $x$
(i.e., $\ell_1\leq \ell \leq \ell_2$).
The algorithm then reads $O(\log n)$ codewords from the compressed string,
using the parent and ancestor pointers in the
codewords, in order to go up the
 trie (implicitly defined by the codewords)
 to the node at distance $\ell_2-\ell$ from the node
 corresponding to $C[t]$.
The final node reached corresponds to the codeword,
$C[r] = (p_r,i_r,k_r,b_r)$, which encodes the
phrase $x[p_r,\dots,\ell-\ell_1+1] = x[\ell_1\dots,\ell]$
and so the algorithm returns $b_r$.

\smallskip\noindent
The next theorem follows directly
from the description of the algorithms and Claim~\ref{clm:span}.
\BT~\label{thm:det}
Algorithm~\ref{alg:com} (compression algorithm) is $3$-competitive with LZ78,
and for every input $x\in\{0,1\}^n$, the running time of
Algorithm~\ref{alg:local} (random access algorithm) is $O(\log n)$.
\ET

\paragraph{Recovering a substring.}
\fmnote{D: elaborate more and/or make formal claim}
We next describe how to recover a consecutive substring
$x[\ell_1, \ldots ,\ell_2]$, given the compressed string $C[1, \ldots, m]$.
The idea is to recover the substring in reverse order as follows.
Find the codeword, $C[k]$ encoding the substring (phrase)
$x[t_1, \ldots, t_2]$ such that $t_1 \leq \ell_2 \leq t_2$ as
\ifnum\dccsub=0
in Step~\ref{local.sp1} of Algorithm~\ref{alg:local}.
\else
described above for the (symbol bit) random access algorithm.
\fi
Then,
\ifnum\dccsub=0
as in Step~\ref{local.sp2} of Algorithm~\ref{alg:local}
\else
as described above as well,
\fi
find the codeword, $C[t]$, which encodes $x[t_1,\dots,\ell_2]$.
From $C[t]$ recover the rest of the substring
($x[t_1,\dots,\ell_2-1]$) by going
up the trie. If the root is reached before recovering
$\ell_2-\ell_1 + 1$ bits (i.e., $\ell_1 < t_1$),
then continue decoding $C[k-1], C[k-2], \ldots$ until
reaching the encoding of the phrase within which $x[\ell_1]$ resides.
The running time is the sum of the running time of
 a single random access execution, plus the length of the
 substring.

\fi
\ifnum\dccsub=0
\section{A Randomized Scheme}\label{random.sec}
In this section we present a randomized compression scheme which
builds on the deterministic scheme described in Section~\ref{sec:det}.
\fi
In what follows we describe the randomized compression algorithm and
the random access algorithm. 
\ifnum\dccsub=0
Their detailed pseudo-codes are given in
Algorithm~\ref{alg:random} (see Subsection~\ref{ranps.subsec}) and
Algorithm~\ref{alg:ran-acc} (see Subsection~\ref{detps.subsec}),
respectively. 
Recovering a substring is done in the same manner as
described for the deterministic scheme.

\else
Their detailed pseudo-codes as well as the formal proof of Theorem~\ref{thm:random} are given in~\cite{DLRR-long}.
\fi
\ifnum\dccsub=0
We assume that $\eps = \Omega(\log n/\sqrt{\log n})$
(or else one might as well  compress using LZ78 without any modifications).
\fi

\ifnum\dccsub=1

{\bf The high-level idea of the compression scheme.}
\else
\paragraph{The high-level idea of the compression scheme.}
\fi
Recall that the deterministic compression
\ifnum\dccsub=0
algorithm (Algorithm~\ref{alg:com}),
\else
algorithm
\fi
which was $3$-competitive, adds to
each LZ78 codeword two additional information entries: the starting position of
the corresponding phrase, and an additional index (pointer) for
navigating up the trie.
The high level idea of the randomized compression algorithm, which
is $(1+\eps)$-competitive, is to ``spread'' this information more
sparsely. That is, rather than maintaining the starting position of
every phrase, it maintains the position only for a
$\Theta(\eps)$-fraction of the phrases, and similarly
only $\Theta(\eps)$-fraction of the nodes in the trie have additional
pointers for ``long jumps''. While spreading out the position
information is done deterministically (by simply adding  this
information once in every $\Theta(1/\eps)$ codewords),
the additional pointers are added randomly (and independently).
Since the trie structure is not known in advance, this ensures
(with high probability) that the number of additional pointer
entries is $O(\eps)$ times the number of nodes (phrases), as
well as ensuring that the additional pointers are fairly
evenly distributed in each path in the trie.
We leave it as an open question whether there exists a deterministic
(online) algorithm that always achieves such a guarantee
\footnote{The simple idea of adding an extra pointer to all the nodes whose depth is divisible by $k = \Theta(1/\eps)$, excluding nodes with height smaller than $k$, will indeed ensure the even distribution on each path.
However, since we do not know the height of each node in advance, if we remove this exclusion we might
cause the number of additional pointers to be too large, e.g., if the trie
is a complete binary tree with height divisible by $k$, then every leaf gets an additional pointer.}.

Because of the sparsity of the position and extra-pointer entries,
finding the exact phrase to which an input bit belongs and navigating
up the trie in order to determine this bit, is not as self-evident
as it was in the deterministic scheme. In particular,
since the position information is added only once every $\Theta(1/\eps)$
phrases,  a binary search (similar to the one performed by the
deterministic algorithm) for a location $\ell$ in the input
string  does not uniquely determine the phrase to which the $\ell$-th
bit belongs. In order to facilitate finding this phrase (among the
$O(1/\eps)$ potential candidates), the compression algorithm adds one
more type of entry to an $O(\eps)$-fraction of the nodes in the trie:
their depth (which equals the length of the phrase to which they correspond).
This information also aids the navigation up the trie, as will be explained
subsequently.

\ifnum\dccsub=1

{\bf A more detailed description of the compression algorithm.}
\else
\paragraph{A more detailed description of the compression algorithm.}
\fi
Similarly to the deterministic compression algorithm,
the randomized compression algorithm
scans the input string and outputs codewords
containing information regarding the corresponding phrases
(where the phrases are the same as defined by LZ78).
However, rather than having just one type of codeword, it
has three types:
\ifnum\dccsub=0
\BI
\item A {\em simple} codeword of the form $(i,b)$,
which is similar to the codeword LZ78 outputs.
Namely, $i$ is a a pointer to a former codeword (which encodes the previously
encountered phrase that is the longest prefix of the current one),
and $b$ is a bit.
Here, since the length of the
codewords is not fixed, the pointer $i$ indicates the starting position
of the former codeword in the compressed string rather than its index.
We refer to $i$ as the {\em parent\/} entry, and to $b$ as the
{\em value\/} entry.

\item A {\em special} codeword, which encodes additional information regarding
the corresponding node in the trie. Specifically, in addition to the
entries $i$ and $b$ as in a simple codeword, there are three additional
entries. One is the {\em depth\/} of the corresponding node, $v$,
in the tree, and the
other two are pointers (starting positions in the compressed string)
to special codewords that correspond to ancestors of $v$.
We refer to one of these entries as the {\em special\_parent\/}
and the other as the {\em special\_ancestor\/}.
Details of how they are selected are given subsequently.
\item
A {\em position} codeword, which contains the starting position of
the next encoded phrase in the uncompressed string.
\EI
\else
(1) A {\em simple} codeword of the form $(i,b)$,
which is similar to the codeword LZ78 outputs.
Namely, $i$ is a a pointer to a former codeword (which encodes the previously
encountered phrase that is the longest prefix of the current one),
and $b$ is a bit.
Here, since the length of the
codewords is not fixed, the pointer $i$ indicates the starting position
of the former codeword in the compressed string rather than its index.
We refer to $i$ as the {\em parent\/} entry, and to $b$ as the
{\em value\/} entry.
(2) A {\em special} codeword, which encodes additional information regarding
the corresponding node in the trie. Specifically, in addition to the
entries $i$ and $b$ as in a simple codeword, there are three additional
entries. One is the {\em depth\/} of the corresponding node, $v$,
in the tree, and the
other two are pointers (starting positions in the compressed string)
to special codewords that correspond to ancestors of $v$.
We refer to one of these entries as the {\em special\_parent\/}
and the other as the {\em special\_ancestor\/}.
Details of how they are selected are given subsequently.
(3) A {\em position} codeword, which contains the starting position of
the next encoded phrase in the uncompressed string.
\fi
In what follows we use the term {\em word\/} (as opposed to {\em codeword\/})
to refer to the RAM words of which the codewords are built.
Since codewords have different types and lengths (in terms of the
number of words they consist of), the compression algorithm
adds a special delimiter word before each special codeword and
(a different one) before each position codeword.\footnote{In particular,
these can be the all-1 word and the word that is all-1 with the exception
of the last bit, which is 0. This is possible because the number
of words in the compressed string is $O(n/\log n)$.}

The algorithm includes a position codeword every $c/\eps$ words
(where $c$ is a fixed constant). More precisely, since such a word
might be in the middle of a codeword, the position codeword is
actually added right before the start of the next codeword
(that is, at most a constant number of words away).
As stated above, the position is the starting position of the phrase
encoded by the next codeword.

Turning to the special codewords, each codeword that encodes a phrase
is selected to be a special codewords independently at random with
probability $\eps/c$. We refer to the nodes in the trie
that correspond to special codewords as {\em special nodes\/}.
Let $u$ be a special node (where this information is maintained
using a Boolean-valued field named `special').
In addition to a pointer $i$ to its parent node in the trie, it
is given a pointer
$q$ to its closest ancestor that is a special node (its
{\em special parent\/}) and a pointer $a$ to a {\em special ancestor\/}.
The latter is determined based on the {\em special depth\/} of $u$,
that is, the number of special ancestors of $u$ plus 1,
similarly to the way it is determined by the deterministic algorithm.
Thus, the special nodes are connected among themselves by a TC-spanner
(with out-degree 2).

\ifnum\dccsub=1

{\bf A more detailed description of the random access algorithm.}
\else
\paragraph{A more detailed description of the random access algorithm.}
\fi
The random access algorithm
\ifnum\dccsub=0
 Algorithm~\ref{alg:ran-acc},
\fi
is given access to a string $S$, which was created by the randomized
\ifnum\dccsub=0
compression algorithm, Algorithm~\ref{alg:random}.
\else
compression algorithm.
\fi
This string consists of codewords $C[1],\dots,C[m]$
(of varying lengths, so that each $C[j]$ equals $S[r,\dots,r+h]$
for $h\in \{0,1,4\}$).
\ifnum\dccsub=0
Similarly to
Algorithm~\ref{alg:local}
for random access when
the string is compressed using the deterministic compression algorithm,
Algorithm~\ref{alg:ran-acc},
the algorithm for random access when
the string is compressed using the randomized compression algorithm,
\else
The algorithm for random access when
the string is compressed using the randomized compression algorithm,
\fi
consists of two stages. Given an index $1 \leq \ell \leq n$,
in the first stage the algorithm finds
the codeword
that encodes the phrase $x[\ell_1,\dots,\ell_2]$
to which the $\ell$-th bit
of the input string $x$ belongs (so that
$\ell_1 \leq \ell \leq \ell_2$). In the second stage it finds
the codeword that encodes the phrase $x[\ell_1,\dots,\ell]$
(which appeared earlier in the string), and returns
its value entry (i.e., the bit $b$).

Recall that on input $\ell$ and $C[1, \ldots, m]$,
\ifnum\dccsub=0
Algorithm~\ref{alg:local} (in Step~\ref{local.sp1})
\else
the random access algorithm for the deterministic scheme
\fi
first finds the codeword 
that encodes the phrase to which the $\ell$-th bit
of the input string  belongs
by performing a binary search. This is done using the position entries,
where each codeword has such an entry.
However,
in the output string of the randomized compression scheme
it is no longer the case that each codeword has a position entry.
Still, the random access algorithm can perform a binary search over the
position codewords. Recall that the randomized compression algorithm
places these codewords at almost fixed positions in the compresses
string (namely, at
positions that are at most a constant number of words away from
the fixed positions), and these codewords are marked by a
delimiter.
\fmnote{D: add numbers of steps in Algorithm~\ref{alg:ran-acc}?}
Hence, the algorithm can find two position codewords,
$C[k]$ and $C[q]$, such that $q-\ell = O(1/\eps)$ and such
that $\ell$ is between the positions corresponding to these
codewords. This implies that the requested bit $x[\ell]$ belongs
to one of the phrases associated with the
codewords $C[k+1],\dots,C[q-1]$.

In order to find the desired codeword $C[t]$ where $k < t < q$,
the algorithm calculates the length of the phrase each
of the codewords $C[k+1],\dots,C[q-1]$ encodes.
This length equals the depth of codeword (corresponding node)
in the trie. If a codeword is a special codeword, then this
information is contained in the codeword.
Otherwise (the codeword is a simple codeword),
the algorithm computes the depth of the corresponding
node by going up the trie until
it reaches a special node (corresponding to a special
codeword). Recall that a walk up the tree can be performed
using the basic parent pointers (contained in both simple
and special codewords), and that each special codeword
is marked by a delimiter, so that it can be easily recognized
as special.
\ifnum\dccsub=0
(For the pseudo-code  see Procedure~\ref{proc:depth} in
Subsection~\ref{ranps.subsec}.)
\fi

Let the phrase encoded by $C[t]$ be $x[\ell_1,\dots,\ell_2]$
(where $\ell_1\leq \ell \leq \ell_2$).
In the second stage, 
the random access algorithm finds the codeword,
$C[r]$, which encodes the phrase $x[\ell_1,\dots,\ell]$
(and returns its value entry, $b$, which equals $x[\ell]$).
This is done in three steps. First the algorithm uses
parent pointers to reach the  special node, $v$, which is closest to
the node corresponding to $C[t]$.
Then the algorithm uses the special\_parent pointers and
special\_ancestor pointers (i.e., TC-spanner edges) to reach
the special node, $v'$, which is closest to the node
corresponding to $C[r]$ (and is a descendent of it).
This step uses the depth information that is provided in
all special nodes in order to avoid ``over-shooting'' $C[r]$.
(Note that the depth of the node corresponding to $C[r]$
is known.) Since the special nodes $v$ and $v'$ are connected
by an $O(\log n)$-TC-spanner, we know (by Claim~\ref{clm:span})
that there is a path of length $O(\log n)$ from $v$ to $v'$.
While the algorithm does not know what is the depth of $v'$,
it can use the depth of the node corresponding to $C[r]$
instead to decide what edges to take.
In the last step, the node corresponding to $C[r]$ is reached
by taking (basic) parent pointers from $v'$.

\ifnum\dccsub=0
\BT~\label{thm:random}
Algorithm~\ref{alg:random} is $(1+\eps)$-competitive with LZ78
and for every input $x\in\{0,1\}^n$, the expected running time of
Algorithm~\ref{alg:ran-acc} is $O(\log n + 1/\epsilon^{2})$.
With high probability over the random coins of
Algorithm~\ref{alg:random} the running time of
Algorithm~\ref{alg:ran-acc} is bounded by
$O(\log n/\eps^2)$.
\ET
\else
\vspace{-1.5ex}
\BT~\label{thm:random}
There exists a compression algorithm which is $(1+\eps)$-competitive with LZ78
and for every input $x\in\{0,1\}^n$, the expected running time of
random access is $O(\log n + 1/\epsilon^{2})$.
With high probability over the random coins of
the compression algorithm the running time of
random access is bounded by
$O(\log n/\eps^2)$.
\ET
\fi

\ifnum\dccsub=0
\BPF 
For an input string $x\in \bitset^n$,
let $w(x)$ be the number of codewords (and hence words)
in the LZ78 compression of $x$, and let $w'(x)$ be the number
of words obtained when compressing with Algorithm~\ref{alg:random}
(so that $w'(x)$ is a random variable). Let $m'_1(x)$ be the
number of simple codewords in the compressed string, let
$m'_2(x)$ be the number of special codewords, and let
$m'_3(x)$ be the number of position codewords.
Therefore, $w'(x) = m'_1(x) + 5m'_2(x) + 2m'_3(x)$.
By construction, $m'_1(x) + m'_2(x) = w(x)$, and
so $w'(x) = w(x) + 4m'_2(x) + 2m'_3(x)$.
Also by construction we have that $m'_3(x) = \eps w'(x)/40$,
so that $w'(x) = \frac{w(x) + 4m'_2(x)}{1-\eps/20}$.
Since each phrase is selected
to be encoded by a special codeword independently with
probability $\eps/40$, by a multiplicative
Chernoff bound, the probability that more than an
$(\eps/20)$-fraction of the phrases will be selected, i.e.,
$m'_2(x) > (\eps/20)w(x)$ is bounded by
$\exp(-\Omega(\eps w(x))) < \exp(-\Omega(\eps\sqrt{n}))$
(since $w(x) \geq \sqrt{n}$).
Therefore, with high probability (recall that
we may assume that $\eps \geq c\log(n)/\sqrt{n}$ for
a sufficiently large constant $c$) we get that
$w'(x) \leq \frac{1+\eps/5}{1-\eps/20}\cdot w(x) \leq (1+\eps)w(x)$.
Since the analysis of the running time is easier to follow
by referring to specific steps in the pseudo-code of the algorithm
(see Subsection~\ref{ranps.subsec})
we refer the reader to
Subsection~\ref{subsec:ranan}
for the rest of the proof.
\EPF 
\fi

\ifnum\dccsub=0
\section{A Lower Bound for Random Access in LZ78}
In what follows we describe a family of strings, $x\in\{0,1\}^n$,
for which random access to $x$ from the LZ78 compressed string,
$C^x = C_{\rm LZ}^x$,
requires $\Omega(|C^x|)$ queries, where $|C^x|$ denotes the number of \
codewords in $C^x$.
We construct the lower bound for strings, $x$, such that  $|C^x| = \Omega(n/\log n)$ (Theorem~\ref{thm:lb}) and afterwards extend (Theorem~\ref{thm:lb2}) the construction for general $n$ and $m$, where $n$ denotes the length of the uncompressed string and $m$ denotes the number
of codewords in the corresponding compressed string.
Note that $m$ is  lower bounded by $\Omega(\sqrt{n})$ and
upper bounded by $O(n/\log n)$. Consider the two extreme cases,
the case where the trie, $T^x$,
has a topology of a line, for example when
$x = \underline{0}\; \underline{01} \; \underline{01^2} \ldots \underline{01^j}$.
In this case $|C^x| = \Omega(\sqrt{n})$; the case where the trie
is a complete tree, corresponding for example to the string that is a
concatenation of all the strings up to a certain length, ordered by their length.
In the latter case, from the fact that $T^x$ is a complete binary tree on $m+1$ nodes it follows that $x$ is of length $\Theta(m\log m)$, thus $|C^x| = O(n/\log n)$.

The idea behind the construction is as follows. Assume $m = 2^k-1$ for some $k\in \mathbb{Z}^+$ and consider
the string $S = \underline{0}\; \underline{1}\; \underline{00}\; \underline{01}\; \underline{10}\; \underline{11}\; \underline{000} \ldots\underline{1^{k-1}}$, namely,
the string that contains all strings of length at most $k-1$ ordered by their length and then by their lexicographical order.
Let $S^{\ell}$ denote the string that is identical to $S$ except for the $\ell$-th order string, $s$, amongst strings with prefix $01$ and length $k-1$. We modify the prefix of $s$ from $01$ to $00$ and add an arbitrary bit to the end of $s$. The key observation is that the encoding of $S$ and $S^{\ell}$ differs in a single location, i.e. a single codeword.
Moreover, this location is disjoint for different values of $\ell$ and therefore implies a lower bound of $\Omega(m)$ as formalized in the next theorem.

\BT\label{thm:lb}
For every $m = 2^k-2$ where $k\in \mathbb{Z}^+$, there exist $n = \Theta(m\log m)$, an index $0 \leq i \leq n$ and a distribution,
$\mathcal{D}$, over $\{0,1\}^n\cup\{0,1\}^{n+1}$ such that
\BE
\item $|C^x| = m$ for every $x\in\mathcal{D}$.\label{con1}
\item Every algorithm $\mathcal{A}$ for which it holds that $\Pr_{x\in \mathcal{D}}\left[\mathcal{A}(C^x) = x_i \right] \geq 2/3$ must read
$\Omega(2^k)$ codewords from $C^x$.\label{con2}
\EE
\ET
\BPF
Let $x\circ y$ denote $x$ concatenated to $y$ and $\bigcirc_{i=1}^t s_i$ denote $s_1\circ s_2 \ldots \circ s_t$. Define $S = \bigcirc_{i=1}^{k-1} \left(\bigcirc_{j=1}^{2^i} s(i, j)\right)$ where $s(i,j)$ is the $j$-th string, according to the lexicographical order, amongst strings of length $i$ over alphabet $\{0,1\}$.
For every $1 \leq \ell \leq q \eqdef 2^{k-1}/4$
define $S^{\ell} = \bigcirc_{i=1}^{k-1}
\left(\bigcirc_{j=1}^{2^i} s^{\ell}(i, j)\right)$ where
$s^{\ell}(i,j) = s(k-1, 1) \circ 0$ for $i=k-1$ and
$j = q + \ell$ and $s^{\ell}(i, j) = s(i, j)$ otherwise.
Define $C^x_{i,j} \eqdef C^x[2^i -1 + j]$.
Therefore, $C^S_{i,j}$ corresponds to the $j$-th node in the $i$-th
level of the $T^S$, i.e. $C^S_{i,j} = s(i,j)$
(see Figure~\ref{fig:lb}, Section~\ref{fig:sec}).
Thus $C^S_{i,j} \neq C^{S^{\ell}}_{i,j}$ for
$\langle i,j\rangle = \langle k-1, q + \ell\rangle$ and
$C^S_{i,j} = C_{i,j}^{S^{\ell}}$ otherwise.
We define $\mathcal{D}$ to be the distribution of the
random variable that takes the value $S$ with probability $1/2$
and the value $S^{\ell}$ with probability $1/(2\ell)$
for every $1 \leq \ell \leq q$.
We first argue that for some absolute constant $\eta< 0$,
for every algorithm, $\mathcal{A}$, which for an input
$C^x$ takes $\eta |C^x|$ queries from $C^x$,
it holds that
$\Pr_{R\in \mathcal{D}}\left[\mathcal{A}(C^S) \neq \mathcal{A}(C^R) \right]
                   \leq 1/6$.
This follows from the combination of the fact that
$q = \Omega(|C^S|)$ and the fact that $\mathcal{A}$ must query the compressed
string
on the $\ell$-th location in order to distinguish $S^{\ell}$ from $S$.
To complete the proof we show that there exists
$0 \leq i \leq n$ such that
$\Pr_{R\in \mathcal{D}}\left[C_i^S = C_i^R \right] = 1/2$,
namely, show that $C^S_i \neq C_i^{S^{\ell}}$ for every
$1 \leq \ell \leq q$. Since the position of the phrases
of length $k-1$ with prefix $1$ is shifted by one
in $S^{\ell}$ with respect to $S$ we get that the above is true for
$\Omega(|C^S|)$ bits. In particular,
$C^S_i \neq C_i^{S^{\ell}}$ holds for every bit, $x_i$,
that is encoded in the second to last position of a phrase
of length $k-1$ with prefix $1$ and suffix $01$.
\EPF

\noindent
Theorem~\ref{thm:lb} can be extended as  follows:
\BT\label{thm:lb2}
For every $\tilde{m}$ and $\tilde{n}$ such that
$\tilde{m}\log \tilde{m} < \tilde{n} < \tilde{m}^2$ there exist:
\BE
\item $m = \Theta(\tilde{m})$ and $n = \Theta(\tilde{n})$
\item a distribution, $\mathcal{D}$, over $\{0,1\}^n\cup\{0,1\}^{n+1}$
\item an index $0 \leq i \leq n$
\EE
such that Conditions~\ref{con1} and~\ref{con2} in Theorem~\ref{thm:lb} hold.
\ET
\BPF
Set $k = \lceil \log \tilde{m}\rceil$, $t = \lceil\sqrt{\tilde{n}}\rceil$ and let $m = 2^k-1 + t$.
Define $S = \bigcirc_{i=1}^{k-1} \left(\bigcirc_{j=1}^{2^i} (0 \circ s(i, j))\right) \bigcirc_{i=1}^{t} 1^i$ and $S^{\ell} = \bigcirc_{i=1}^{k-1} \left(\bigcirc_{j=1}^{2^i} (0 \circ s^{\ell}(i, j))\right) \bigcirc_{i=1}^{t} 1^i$.
Therefore $n = \Theta(k2^k  + t^2) = \Theta(\tilde{n})$.
The rest of the proof follows the same lines as in the proof of Theorem~\ref{thm:lb}.
\EPF
\fi
\section{Experimental Results}
\begin{figure}[th]
\begin{tikzpicture}
	\begin{axis}[
	xlabel=$\alpha$ - Fraction of Special Codewords,
	ylabel=File Size (KB),
		height=7cm,
		width=9cm,
		grid=major,
legend style={at={( 1.1,1)}, anchor=north west}
	]
		
	\addplot coordinates {
		(0,492)
		(0.0625,586)
                     (0.125,684)
                     (0.25,885)
	};
	\addlegendentry{log file 1.2MB}

           \addplot coordinates {
		(0,594)
(0.0625,705)
(0.125,823)		
(0.25,1126.4)		
	};
	\addlegendentry{DNA sequence 1.7MB}
	
         \addplot coordinates {
		(0,655)
(0.0625,782)		
		(0.125,913)
(0.25,1228.8)		
	};
	\addlegendentry{vocabulary 1.3MB}

	\end{axis}
\end{tikzpicture}
\caption{Competitive Ratio}\label{com.fig}
\end{figure}
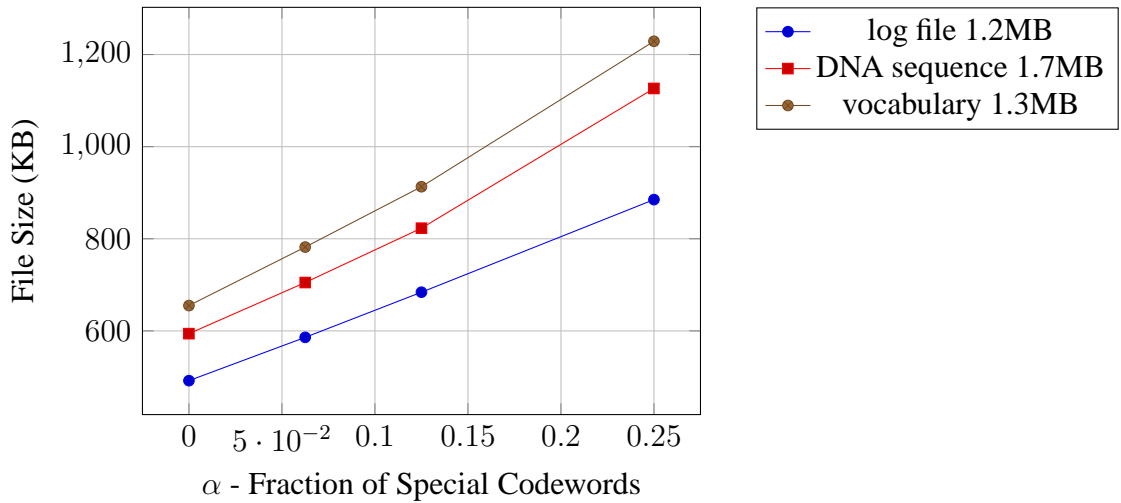

Our experiments show that on selected example files our scheme is competitive in practice (see Figure~\ref{com.fig}).
Our results are given below in terms of the fraction of special codewords, $\alpha$, which is directly related to $\eps$ (see Theorem~\ref{thm:random}). 
We ran the scheme with $\alpha = 1/4, 1/8, 1/16$. The data points corresponding to $\alpha = 0$  plot the file size resulting from standard LZ78.

With respect to the random access efficiency, we found that on average the time required for random access is less than $1$ millisecond while decompressing the entire file takes around $300$ milliseconds.    
\paragraph{Acknowledgment.}
We would like to thank Giuseppe Ottaviano for a helpful discussion and two anonymous referees for their constructive comments.

\ifnum\dccsub=1
\vspace{-1.5ex}
\fi
\begin{spacing}{0.0}
\bibliographystyle{plain}
\bibliography{refs}
\end{spacing}
\ifnum\dccsub=0
\newpage
\appendix

\section{Pseudo-code}\label{psdo:sec}
\subsection{Deterministic Scheme Pseudo-code}\label{detps.subsec}
\begin{algorithm}[ht!]
\caption{Deterministic Compression Algorithm}
\label{alg:com}
\KwIn{$x\in \{0,1\}^n$}
Initialize $\mathcal{T}$ to a single root node;
$p := 1$, $j := 1$.\\
While ($p \leq n$)
\BE
\item Find a path in $\mathcal{T}$ from the root to the
deepest node, $v$,
which corresponds to a prefix of $x[p,\ldots, n-1]$, where $0$ corresponds
to the left child and $1$ corresponds to the right child.
\item Create a new node $u$ and set $u.{\rm index} := j$,
$u.{\rm depth} := v.{\rm depth} +1$.
\item If $x[p+v.{\rm depth}]=0$ set $v.{\rm left} := u$
and otherwise set $v.{\rm right} := u$.
\item\label{st:anc} Let $a$ be the ancestor of $u$ in $\mathcal{T}$
at depth $\max \{u.{\rm depth} - 2^{f_n(u.{\rm depth})}\cdot \bsize, 0\}$.
\item\label{st:com-out}
 Output $(p, v.{\rm index}, a.{\rm index}, x[p+u.{\rm depth}])$.
\item $p := p + u.{\rm depth}+1$.
\item $j := j +1$.
\EE
\end{algorithm}

\begin{algorithm}[th]\label{alg:local}
\caption{Random Access Algorithm for Deterministic Scheme}
\KwIn{$C[1]=(p_1,i_1,k_1,b_1), \ldots C[m]=(p_m, i_m, k_m,b_m)$, which
represents a string compressed by Algorithm~\ref{alg:com},
and an index $1\leq \ell \leq n$}
\BE
\item Perform a binary search on $p_1, \ldots, p_m$ and find $p_t$ such that
$p_t = \max_{1\leq i\leq m}\{p_i \leq \ell\}$.\label{local.sp1}
\item Find the codeword, $C[r] = (p_r, i_r, k_r, b_r)$, which correspond to the
 ancestor of $C[t] = (p_t, i_t, k_t,b_t)$ at depth $\ell-p_t+1$
 in the trie. This is done
as described in the proof of Claim~\ref{clm:span}
using the pointer information in the codewords/nodes
(observe that the depth of $C[t]$ is $p_{t+1}-p_t$).
\label{local.sp2}
\item Output $b_r$.\label{local.sp3}
\EE
\end{algorithm}

\clearpage
\subsection{Randomized Scheme Pseudo-code}\label{ranps.subsec}

\begin{algorithm}[ht!]
\label{alg:random}
\caption{Randomized Compression Algorithm}
\KwIn{$x\in \{0,1\}^n$, $\epsilon$}
Initialize $T$ to a root node, $p := 1$, $j := 1$\\
While ($p \leq n$)
\BE
\item Find a path in $\mathcal{T}$ from the root to a leaf, $v$,
which corresponds to a prefix of $x[p,\ldots, n]$, where $0$
corresponds to left child and $1$ corresponds to right child.
\item Create a new node $u$ and set:
\BI
\item $u.{\rm index} := j$
\item $u.{\rm depth} := v.{\rm depth} +1$
\item $u.{\rm special} := 0$
\EI
\item If $x[p+v.{\rm depth}]=0$ set $v.{\rm left} := u$
and otherwise set $v.{\rm right} := u$.
\item $h := j\mod 40/\epsilon$.
\item Toss a coin $c$, with success probability $\epsilon/40$.
\item If $c=1$ output a special codeword as follows:
\BE
\item $u.{\rm special} := 1$
\item Let $\mathcal{P}$ denote the path in $T$ from $u$ to the root and
let $q$ be the first node in $\mathcal{P}$ such that $q.{\rm special} = 1$
(if such exists, otherwise $q=0$).
\item If $q \neq 0$ set
$u.{\rm special\_depth} := q.{\rm special\_depth} + 1$,
otherwise $u.{\rm special\_depth} := 0$.
\item Let $d := u.{\rm special\_depth}$. If $d\neq 0$, let
$a$ be the special node on $\mathcal{P}$ for which
$a.{\rm special\_depth} = \max\left\{d-2^{f_n(d)}\cdot \bsize,0\right\}$.
\item $j := j + 4$
\item Output
$(\bigtriangleup, u.{\rm depth}, v.{\rm index}, q.{\rm index},
a.{\rm index}, x[p+u.{\rm depth}])$,
($\bigtriangleup$ is a delimiter symbol)
\EE
Else, output a simple codeword, namely $i, x[p+u.{\rm depth}]$.
\item $p := p + u.{\rm depth} + 1$.
\item $j := j +1$.
\item \label{st1.1}If $h > (j\mod 40/\epsilon)$, output
$\bigtriangledown, p$ ($\bigtriangledown$ is a delimiter symbol)
\EE
\end{algorithm}

\begin{algorithm}
\caption{Random Access Algorithm for Randomized Scheme}
\label{alg:ran-acc}
\KwIn{a string, $S$, which is the output Algorithm~\ref{alg:random},
and an index $1\leq \ell \leq n$. $S$ consists of varying length codewords
$C[1],\dots,C[m]$}
\BE
\item\label{st1}\label{ran-acc.sp1}
Perform a binary search on the position codewords in
 $S$ to find a position codeword
$C[k]$ such that $C[k].{\rm position}\leq \ell$ and $C[q].{\rm position} > \ell$
where $C[q]$ is the next position codeword in $S$.
\item $p := C[k].{\rm position}$
\item \label{st2}\label{ran-acc.sp2}
Starting from $C[k+1]$,
scan $S$ and find the codeword, $C[t]$, which encodes
the phrase that contains the bit at position $\ell$ as follows:
\BE
\item $t := k+1$
\item $d :=$~\ref{proc:depth}$(C[t])$
\item While ($p + d < \ell$)
\BE
\item $p := p + d$
\item Read the next codeword, $C[t]$.
\item $d :=$~\ref{proc:depth}$(C[t])$
\EE
\EE
\item \label{ran-acc.sp3}
$C[r] :=$~\ref{proc:find}$(C[t], \ell - p + 1)$
\item Output $C[r].{\rm value}$
\EE
\end{algorithm}

\begin{figure*}[ht!]
\begin{minipage}[t]{3.15in}
\begin{procedure}[H]
\caption{Find-Node-by-Depth($u, d$)}
\label{proc:find}
\KwIn{the source node, $u$, and the depth of the target node, $d$}
\BE
\item\label{st:call-proc-depth}
 $s:=$~\ref{proc:depth}$(u) - d$
\item\label{whl1} While ($u$ is not a special node and $s > 0$)
\BE
\item $u := u.{\rm parent}$
\item $s := s - 1$
\EE
\item $v := u.{\rm special\_parent}$
\item\label{whl2}
While $v.{\rm special\_ancestor.depth} < u.{\rm special\_ancestor.depth}$
\BE
\item If ($v.{\rm special\_parent.depth} < d$) then break loop
\item Else, $u := v$
\EE
\item\label{whl3} While ($u.{\rm special\_parent.depth} \geq d$)
\BE
\item If ($u.{\rm special\_ancestor.depth} \geq d$)
then $u := u.{\rm  special\_ancestor}$
\item Else, $u := u.{\rm special\_parent}$
\EE
\item $s := u.{\rm depth} - d$\\
\item\label{whl4} While ($s > 0$)
\BE
\item $u := u.{\rm parent}$
\item $s := s - 1$
\EE
\item Output $u$
\EE
\end{procedure}
\end{minipage}
 \hfill
 \begin{minipage}[t]{3in}
\begin{procedure}[H]
\caption{Find-Depth($u$)}
\label{proc:depth}
\KwIn{source node $u$}
If $u$ is a special node, return $u.{\rm depth}$.
$i := 1$\\
While($u.{\rm parent}$ is a simple node)
\BE
\item $u := u.{\rm parent}$
\item $i := i + 1$
\EE
Return $i + u.{\rm depth}$
\end{procedure}
 \end{minipage}
 \hfill
\end{figure*}


\clearpage
\section{Running Time Analysis and Improvement}
\subsection{Bounding the Running Time of Algorithm~\ref{alg:ran-acc}}\label{subsec:ranan}
In Step~\ref{st1}, Algorithm~\ref{alg:ran-acc} performs a binary search,
therefore it terminates after at most $\log n$ iterations.
In each iteration of the binary search the algorithm scans a constant
number of words as guaranteed by Step~\ref{st1.1} in
Algorithm~\ref{alg:random}.
Hence, the running time of Step~\ref{st1} is bounded by $O(\log n)$.

In order to analyze the remaining steps in the algorithm, consider
any node $v$ in $T$.
Since each node is  picked
to be special with probability $\eps/40$, the expected distance
of any node to the closest special node is $O(1/\eps)$.
Since the choice of special nodes is done independently,
the probability that
the closest special ancestor is at distance greater than
${40c\log n}/{\eps}$ is $(1-\eps/40)^{40c\log n/\eps} < 1/n^c$.
By taking a union bound over all $O(n)$ nodes, with high probability,
 for every node $v$ the closest special
ancestor is at distance $O(\log n/\eps)$.

The first implication of the above is that the running
time of Procedure~\ref{proc:depth} is $O(1/\eps)$ in expectation,
and with high probability every call to Procedure~\ref{proc:depth}
takes time $O(\log n/\eps)$. Hence Step~\ref{ran-acc.sp2} in
Algorithm~\ref{alg:ran-acc} takes time $O(1/\eps^2)$ in expectation
and $O(\log n/\eps^2)$ with high probability.
It remains to upper bound the running time of Procedure~\ref{proc:find} (see Subsection~\ref{ranps.subsec}),
which is called in Step~\ref{ran-acc.sp3} of Algorithm~\ref{alg:ran-acc}.

With high probability, the
running time of Steps~\ref{st:call-proc-depth},~\ref{whl1} and~\ref{whl4}
in Procedure~\ref{proc:find} is $O(1/\eps)$ in expectation, and
$O(\log n/\eps)$ with high probability. The running time of
Step~\ref{whl2} is $O(\log n)$ be the definition of the TC-spanner
over the special nodes.
Finally, by
the explanation following the description of the algorithm
regarding the relation between Step~\ref{whl3}
in Procedure~\ref{proc:find} and the path constructed in
the proof of Claim~\ref{clm:span}, the running time
of Step~\ref{whl3} is $O(\log n)$ as well.
Summing up all contribution to the running time we get the
bounds stated in the lemma.

\subsection{Improving the Running Time from $O(\log n/\eps^2)$ to $\tilde{O}(\log n/\eps + 1/\eps^2)$}\label{sec:imp-run}
As can be seen from the proof of Theorem~\ref{thm:random},
the dominant contribution to the running time of the random access
algorithm (Algorithm~\ref{alg:ran-acc}) in the worst case
(which holds with high probability)
comes from Step~\ref{ran-acc.sp2} of the algorithm.
We bounded the running time of this step by $O(\log n/\eps^2)$
while the running time of the others steps is bounded by
$O(\log n/\eps)$.
In this step the algorithm computes the length of $O(1/\eps)$
phrases by determining the depth in the trie of their corresponding
nodes. This is done by walking up the trie until a special
node is reached. Since we bounded (with high probability)
 the distance of every node to the closest special node
 by $O(\log n/\eps)$, we got $O(\log n /\eps^2)$.
 However, by modifying the algorithm and the analysis, we can
 decrease this bound to $\tilde{O}(\log n/\eps + 1/\eps^2)$.
 Since this modification makes the algorithm a bit more complicated,
 we only sketch it below.

Let $v_1,\dots,v_k$, where $k = O(1/\eps)$ be the nodes
whose depth we are interested in finding. Let $T'$ be the tree
that contains all these nodes and their ancestors in the trie.
Recall that the structure of the trie is determined by the
LZ78 parsing rule, which is used by our compression algorithm,
and that the randomization of the algorithm is only over the
choices of the special nodes.
To gain intuition, consider two extreme cases. In one case
$T'$ consists of a long path, at the bottom of which is a
complete binary tree, whose nodes are $v_1,\dots,v_k$. In
the other extreme, the least common ancestor of any two nodes
$v_i$ and $v_j$ among $v_1,\dots,v_k$, is very far away from
both $v_i$ and $v_j$. Consider the second case first, and let
$X_1,\dots,X_k$ be random variables whose value is determined by
the choice of the special nodes in $T'$, where $X_i$ is the
distance from $v_i$ to its closest ancestor that is a special node.
In this (second) case $X_1,\dots,X_k$ are almost independent.
Assuming they were truly independent, it is not hard to
show that with high probability (i.e., $1-1/\poly(n)$),
not only is each $X_i$ upper bounded by $O(\log n/\eps)$,
but so is there sum. Such a bound on the sum of the $X_i$'s
directly gives a bound on the running time of
Step~\ref{ran-acc.sp2}.

In general, these random variables may be very dependent.
In particular this is true in the first aforementioned case.
However, in this (first) case, even if none of the nodes
in the small complete tree are special, and the distance
from the root of this tree to the closest special node is
$\Theta(\log n/\eps)$, we can find the depth of
all nodes $v_1,\dots,v_k$ in time $O(\log n/\eps)$ (even though
the sum of their distances to the closest special node is
$O(\log n/\eps^2)$. This
is true because once we find the depth of one node by walking
up to the closest special node, if we maintain
the information regarding the nodes passed on the way,
we do not have to take the same path up $T'$ more than once.
Maintaining this information can be done using standard
data structures at a cost of $O(\log(\log n/\eps))$
per operation. As for the analysis, suppose we redefine
$X_i$ to be the number of steps taken up the trie until
either a special node is reached, or another node whose
depth was already computed is reached. We are interested
in upper bounding $\sum_{i=1}^k X_i$. Since these random
variables are not independent, we define a set of i.i.d.
random variables, $Y_1,\dots,Y_k$, where each $Y_i$ is
the number of coins flipped until a `HEADS'  is obtained,
where each coin has bias $\eps/c$. It can be verified
that by showing that with high probability
$\sum_{i=1}^k Y_i = O(\log n/\eps)$ we can get the
same bound for $\sum_{i=1}^k X_i$, and such a bound
can be obtained by applying a multiplicative Chernoff bound.

\section{Figures}\label{fig:sec}

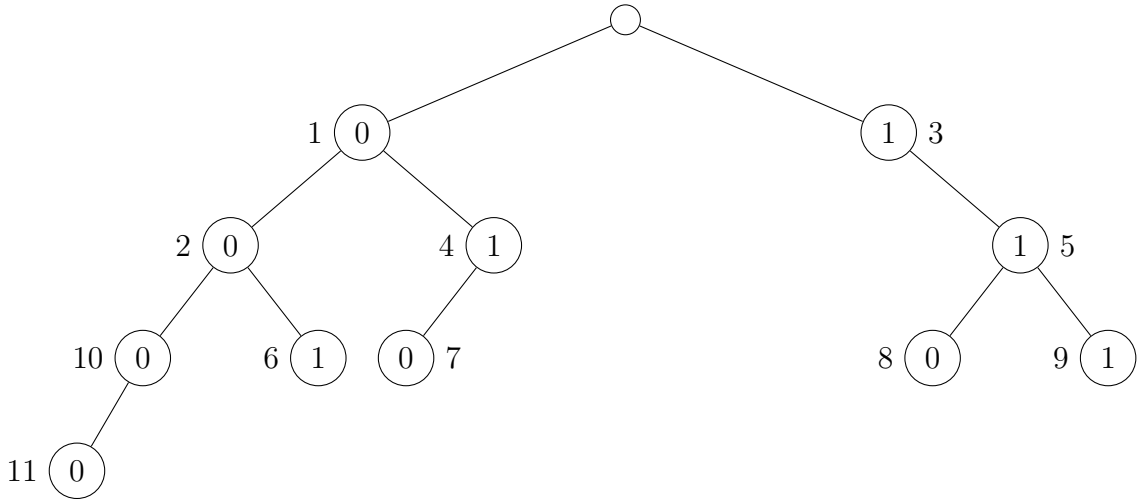
\begin{figure}[ht!]
\ifnum\nofigures=0
\centering
\begin{tikzpicture}[level/.style={sibling distance=70mm/#1}]
\node [circle,draw] (z){}
  child {node [circle,draw,label=left:{$1$}] (a) {$0$}
    child {node [circle,draw,label=left:{$2$}] (b) {$0$}
        child {node [circle,draw,label=left:{$10$}] (c) {$0$}
            child {node [circle,draw,label=left:{$11$}] (g) {$0$}}
            child{edge from parent[draw=none]} }
        child {node [circle,draw,label=left:{$6$}] (d) {$1$}}}
    child {node [circle,draw,label=left:{$4$}] (e) {$1$}
        child {node [circle,draw,label=right:{$7$}] (f) {$0$}}
        child{edge from parent[draw=none]}}}
  child {node [circle,draw,label=right:{$3$}] (h) {$1$}
    child {edge from parent[draw=none]}
    child {node [circle,draw,label=right:{$5$}] (l) {$1$}
        child {node [circle,draw,label=left:{$8$}] (m) {$0$}}
        child {node [circle,draw,label=left:{$9$}] (n) {$1$}}}}
;
\end{tikzpicture}
\fi
\caption{The trie, $T^x$, implicitly defined by the LZ78 scheme on
the string $x = \underline{0}\;\underline{00}\; \underline{1}\; \underline{01}\;
  \underline{11}\; \underline{001}\; \underline{010}\; \underline{110}\;
  \underline{111}\; \underline{000}\; \underline{0000}$;
  On input string $x$, the LZ78 scheme outputs a list of codewords,
  $C^x = \{(0, 0), (1,0), (0, 1), (1,1), (3,1), (2,1), (4,0), (5,0), (5,1),
     (2,0),(10,0)\}$.}
\label{fig:lz}
\end{figure}

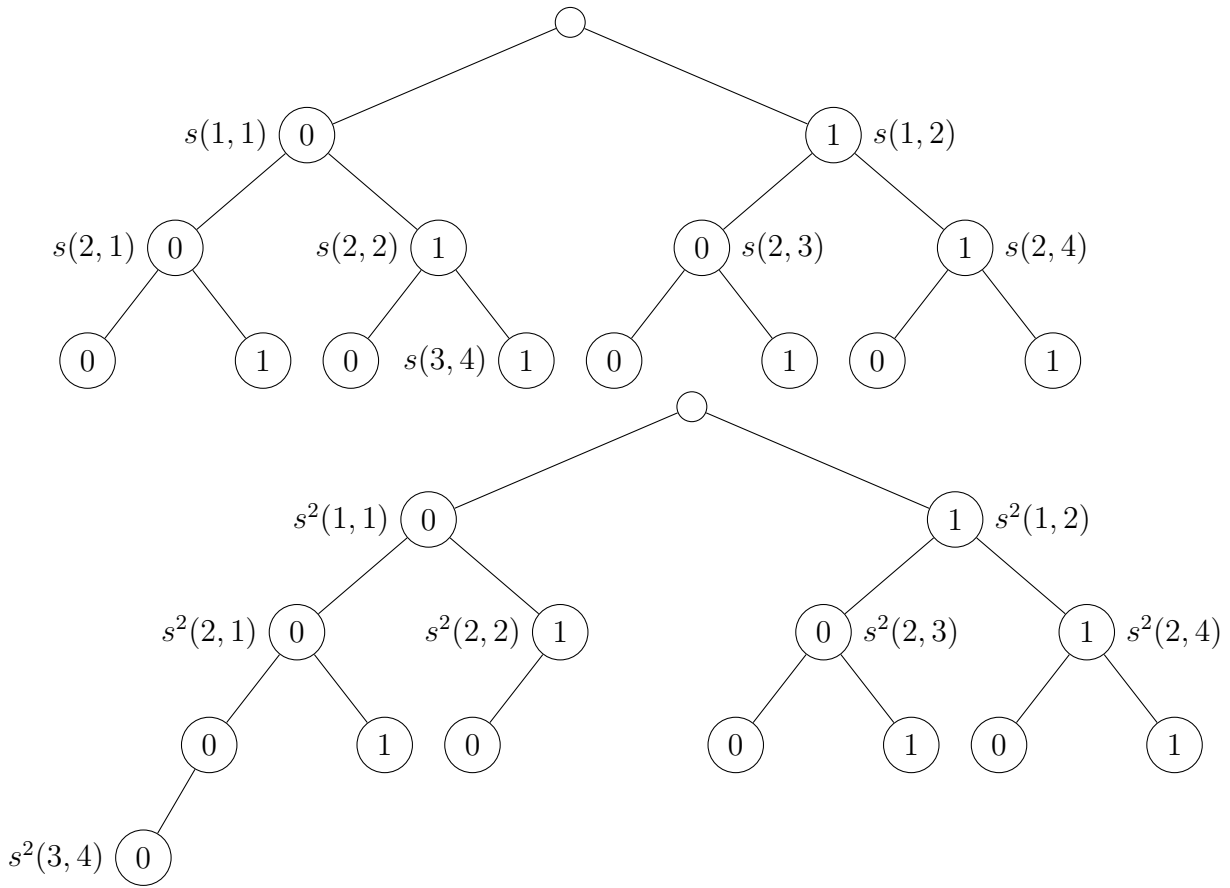
\begin{figure}[ht!]
\ifnum\nofigures=0
\centering
\begin{tikzpicture}[level/.style={sibling distance=70mm/#1}]
\node [circle,draw] (z){}
  child {node [circle,draw,label=left:{$s(1,1)$}] (a) {$0$}
    child {node [circle,draw,label=left:{$s(2,1)$}] (b) {$0$}
        child {node [circle,draw] (c) {$0$}}
        child {node [circle,draw] (d) {$1$}}}
    child {node [circle,draw,label=left:{$s(2,2)$}] (e) {$1$}
        child {node [circle,draw] (f) {$0$}}
        child {node [circle,draw,label=left:{$s(3,4)$}] (g) {$1$}}}}
  child {node [circle,draw,label=right:{$s(1,2)$}] (h) {$1$}
    child {node [circle,draw,label=right:{$s(2,3)$}] (i) {$0$}
        child {node [circle,draw] (j) {$0$}}
        child {node [circle,draw] (k) {$1$}}}
    child {node [circle,draw,label=right:{$s(2,4)$}] (l) {$1$}
        child {node [circle,draw] (m) {$0$}}
        child {node [circle,draw] (n) {$1$}}}}
;
\end{tikzpicture}
\begin{tikzpicture}[level/.style={sibling distance=70mm/#1}]
\node [circle,draw] (z){}
  child {node [circle,draw,label=left:{$s^2(1,1)$}] (a) {$0$}
    child {node [circle,draw,label=left:{$s^2(2,1)$}] (b) {$0$}
        child {node [circle,draw] (c) {$0$}
            child {node [circle,draw,label=left:{$s^2(3,4)$}] (g) {$0$}}
            child{edge from parent[draw=none]} }
        child {node [circle,draw] (d) {$1$}}}
    child {node [circle,draw,label=left:{$s^2(2,2)$}] (e) {$1$}
        child {node [circle,draw] (f) {$0$}}
        child{edge from parent[draw=none]}}}
  child {node [circle,draw,label=right:{$s^2(1,2)$}] (h) {$1$}
    child {node [circle,draw,label=right:{$s^2(2,3)$}] (i) {$0$}
        child {node [circle,draw] (j) {$0$}}
        child {node [circle,draw] (k) {$1$}}}
    child {node [circle,draw,label=right:{$s^2(2,4)$}] (l) {$1$}
        child {node [circle,draw] (m) {$0$}}
        child {node [circle,draw] (n) {$1$}}}}
;
\end{tikzpicture}
\fi
\caption{$T(S)$ and $T(S^2)$}\label{fig:lb}
\end{figure}
\fi

\end{document}